\documentclass[12pt]{iopart}

\usepackage{iopams}

\usepackage{graphicx}
\usepackage{amssymb}
\usepackage{epsfig}

\def\be{\begin{equation}}
\def\ee{\end{equation}}

\begin{document}

\title{QCD phase diagram and charge fluctuations}

\author{K. Redlich$^{a,b}$, B. Friman$^b$ and C. Sasaki$^b$}

\address{
$^a$ Institute of Theoretical Physics, University of Wroclaw, PL--50204 Wroc\l aw, Poland
\\
$^b$ Gesellschaft f\"ur Schwerionenforschung, GSI, D-64291 Darmstadt, Germany}

\begin{abstract}
We discuss the phase structure and fluctuations of conserved charges  in two flavor QCD.
The importance   of the  density fluctuations  to probe  the existence of the critical
end point is summarized. The role of these fluctuations  to identify the first order
phase transition in the presence of spinodal phase separation is also discussed.
\end{abstract}

\vspace{2mm}

\section{Introduction}

One of the essential predictions of QCD is the existence of a phase boundary in the
$(T,\mu_q)$--plane  that separates the chirally broken  hadronic phase from  chirally
symmetric quark-gluon plasma phase \cite{st1}. Arguments based on effective model
calculations [1-12]
%
%
 indicate that at large
$\mu_q$ the transition along this boundary line can  be  first order. In the opposite
limit of high temperature and  low baryon number density the transition from hadronic to
quark gluon plasma phase is expected to be continuous or second order depending  on the
number of quark flavors and the quark masses [7-12].
 This  suggests that the QCD phase diagram  can exhibit a critical endpoint where  the line
of first order phase transitions matches that of second order or analytical crossover
\cite{stephanov}. The critical properties of this second order critical endpoint are
expected to be determined by the Ising model universality class \cite{ef5,hatta,ising}.
The existence of  a critical end point in QCD has   recently been studied in lattice
calculations at non-vanishing chemical potential in 2- and (2+1)--flavor QCD
\cite{LGT1,LGT2,LGT3}.

In this paper we discuss how a  generic QCD phase diagram appears in effective chiral
model calculations. We  consider  the properties of the fluctuations of the conserved
charges to identify the position of the phase boundary and  the critical end point.
 We
compare the model results with  recent LGT findings in 2-flavor QCD and discuss their
physical interpretation.  Finally,  considering deviations from an idealized equilibrium
picture of the first order chiral phase transition  we discuss the properties of charge
density fluctuations by including spinodal instabilities.
\begin{figure}
%
{\includegraphics[width=7.4cm,height=4.7cm]{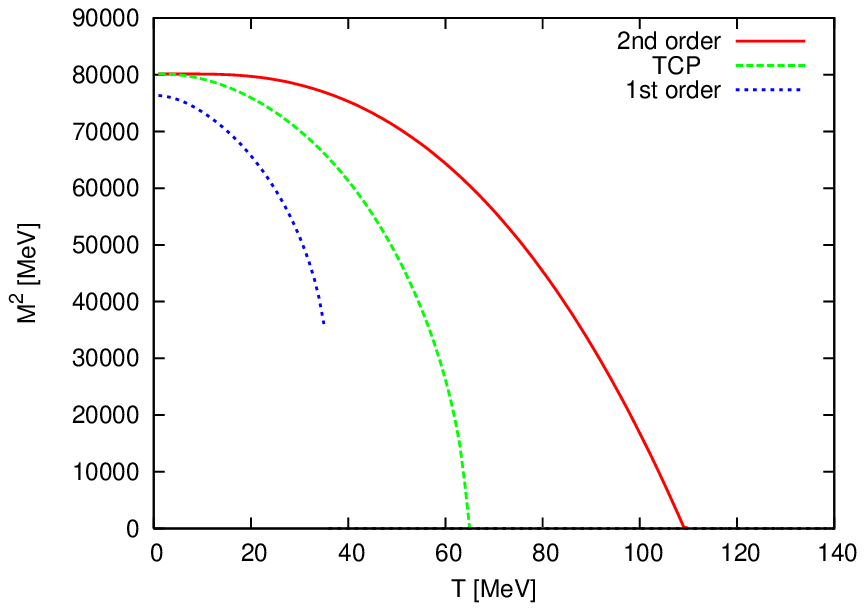}} {\vskip -11.35pc
\hskip 19.3pc
\includegraphics[width=7.4cm,height=4.9cm]{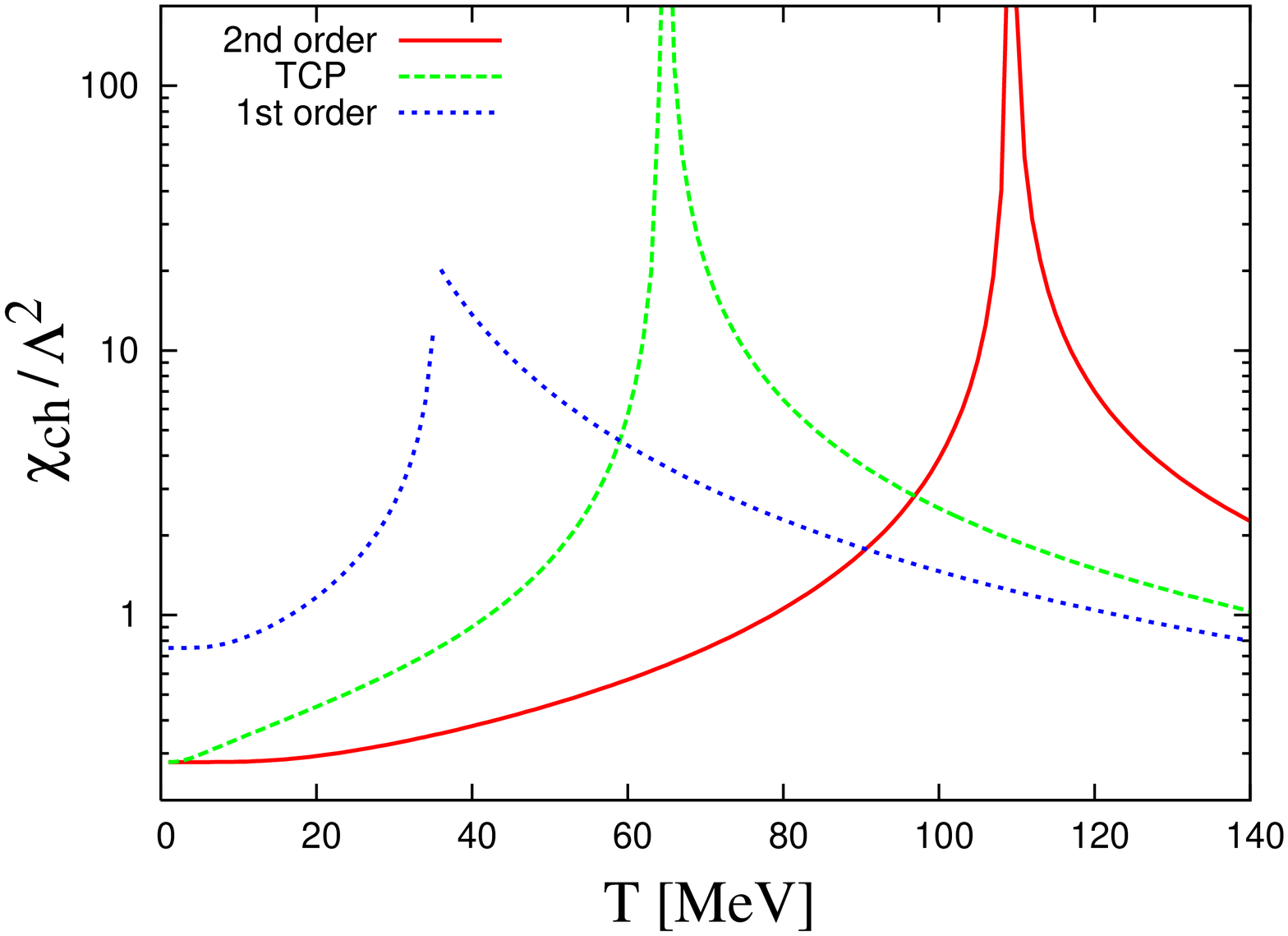}}
\begin{center}\vskip -0.3cm
 \caption{\label{figphase} The chiral condensate $M$ (left-hand figure) and the chiral
susceptibility $\chi_{ch}$ (right-hand figure) calculated  in the NJL model in the chiral
limit as  a function of temperature  for different values of $\mu_q=300, 275$ and 200 MeV
  corresponding  to a 1st order, the tricritical  point and a 2nd order transition
\protect{\cite{our}} . }
\end{center}
\vspace*{0.5cm}
\end{figure}
\section{Charge density fluctuations in  effective chiral models and in LGT}

 To study  thermodynamics and the phase structure related with the chiral symmetry
  in QCD we adopt the  Nambu--Jona-Lasinio model formulated
in the mean field approximation \cite{klevansky,our}. In the chiral limit this  model
describes the  chiral phase transition from  a phase of massive quarks with dynamically
generated mass $M(T,\mu_q)$ dependent  on temperature $T$ and quark chemical potential
$\mu_q$ to a phase of massless quarks.


The effective quark mass  $M(T,\mu_q)$ is a measure of  the thermal expectation value of
the quark condensate at finite   $T$ and $\mu_q$. Thus, $M(T,\mu_q)$ acts as an order
parameter for chiral symmetry restoration. In the chirally broken phase $M\neq 0$ and it
vanishes when the symmetry is restored. In the Fig. 1-left we show the  $T$-dependence of
the dynamical quark mass for different values of $\mu_q$. For small $\mu_q$ the
condensate continuously melts in the narrow temperature range   indicating the second
order phase transition. At large $\mu_q$ the condensate $M$ drops discontinuously from a
finite to zero value indicating the first order nature of the chiral transition. The
fluctuations of the order parameter $\chi_{ch}=<M^2>-<M>^2$, shown in the Fig. 1-right,
are finite when crossing the first order transition whereas they   diverge at the second
order line. The divergence is directly linked  to the appearance of a massless mode in
the scalar-isoscalar, so called   sigma channel,   if the transition is  second order.
The resulting phase diagram obtained in the NJL model with the particular set of the
model parameters and in the chiral limit is shown in the Fig.2-left. It exhibits the
generic, QCD-like structure discussed in the introduction. For a finite quark mass the
second order transition is converted to a  cross-over transition \cite{st1}.

\begin{figure}
%
\hskip 0.5cm {\includegraphics[width=7.1cm,height=5.1cm]{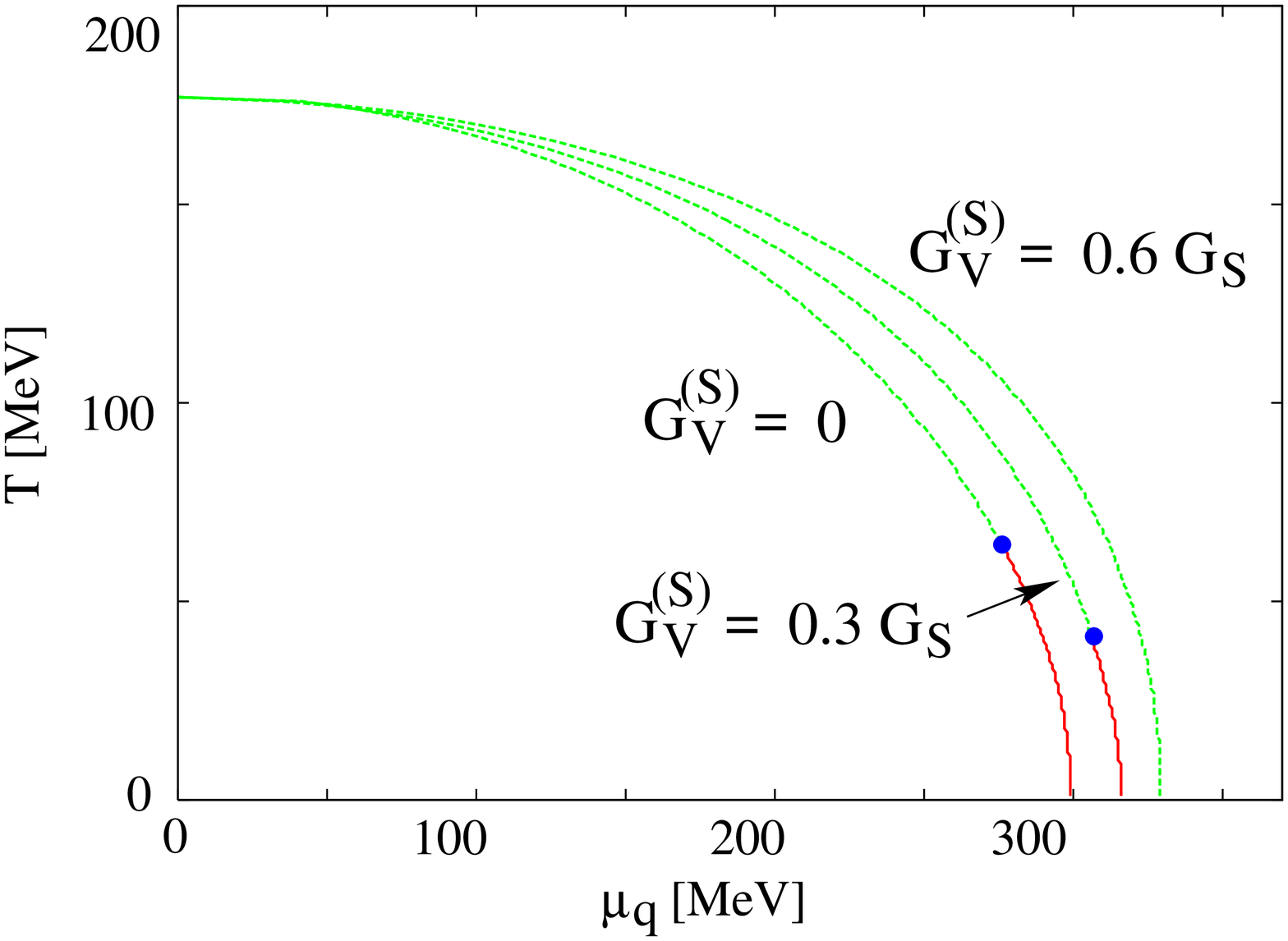}} {\vskip
-12.35pc
\hskip 19.3pc
\includegraphics[width=7.1cm,height=5.1cm]{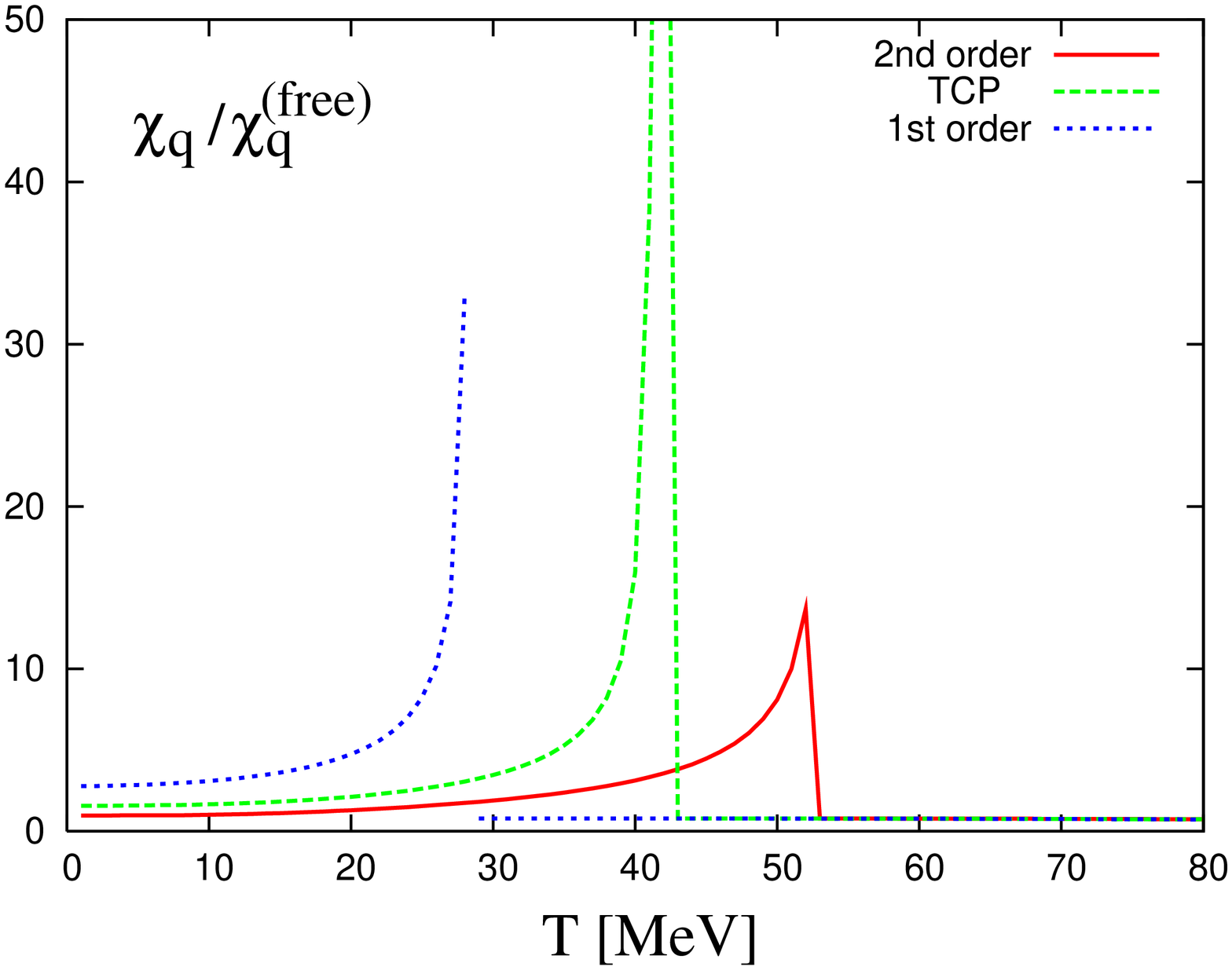}}
\begin{center}\vskip -0.3cm
\caption{\label{fig1:phase} The left-hand figure: The NJL model phase diagram in the
chiral limit for different values of the isovector-vector coupling  $G_V$. The right-hand
figure shows  the net quark number  fluctuations for $\mu_q=310,305$ and 300 MeV
{\cite{our}}. }
\end{center}
\vspace*{0.5cm}
\end{figure}
\begin{figure}
%
\hskip 0.5cm {\includegraphics[width=7.2cm,height=5.4cm]{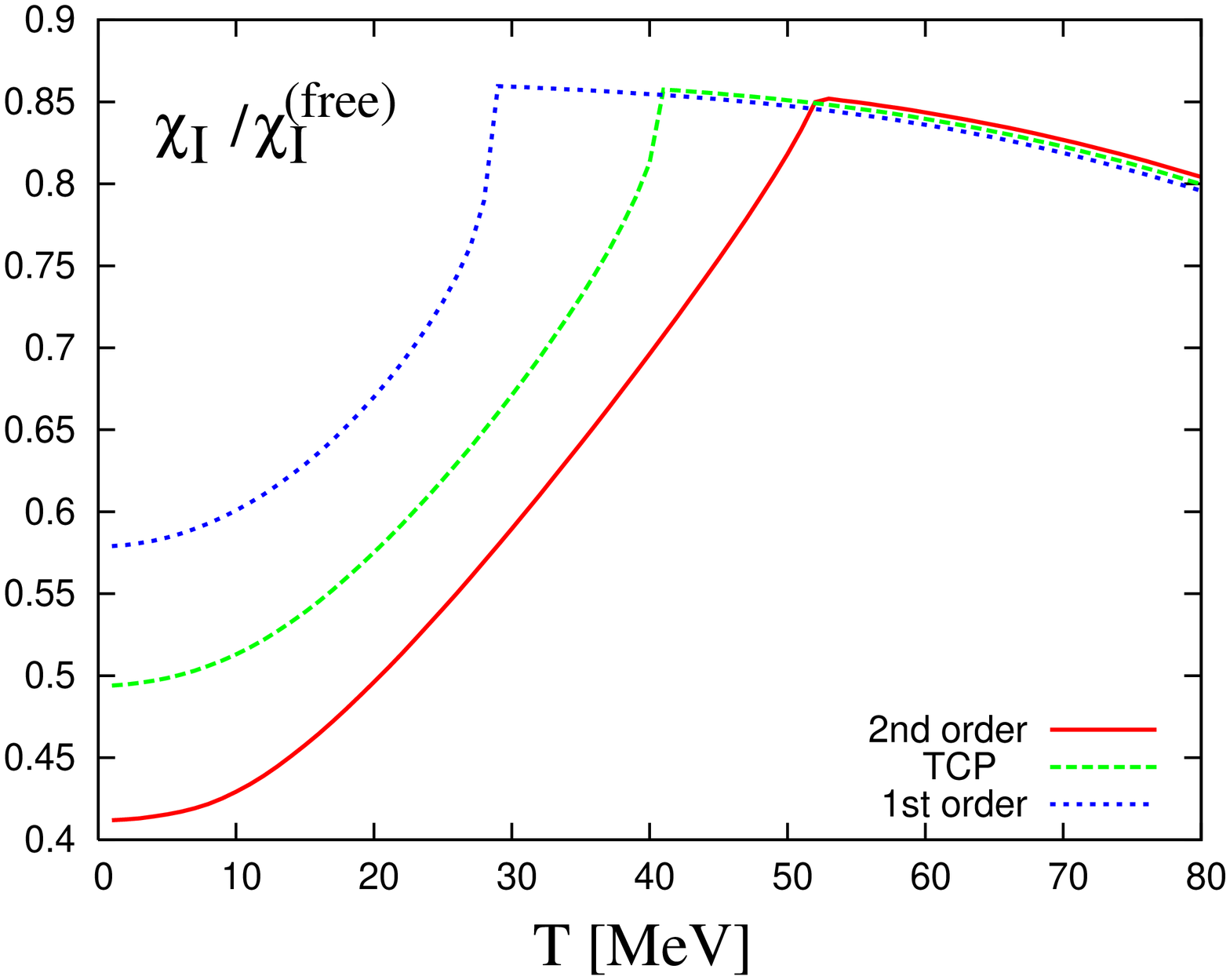}} {\vskip -12.95pc
\hskip 18.pc
\includegraphics[width=7.6cm,height=5.5cm]{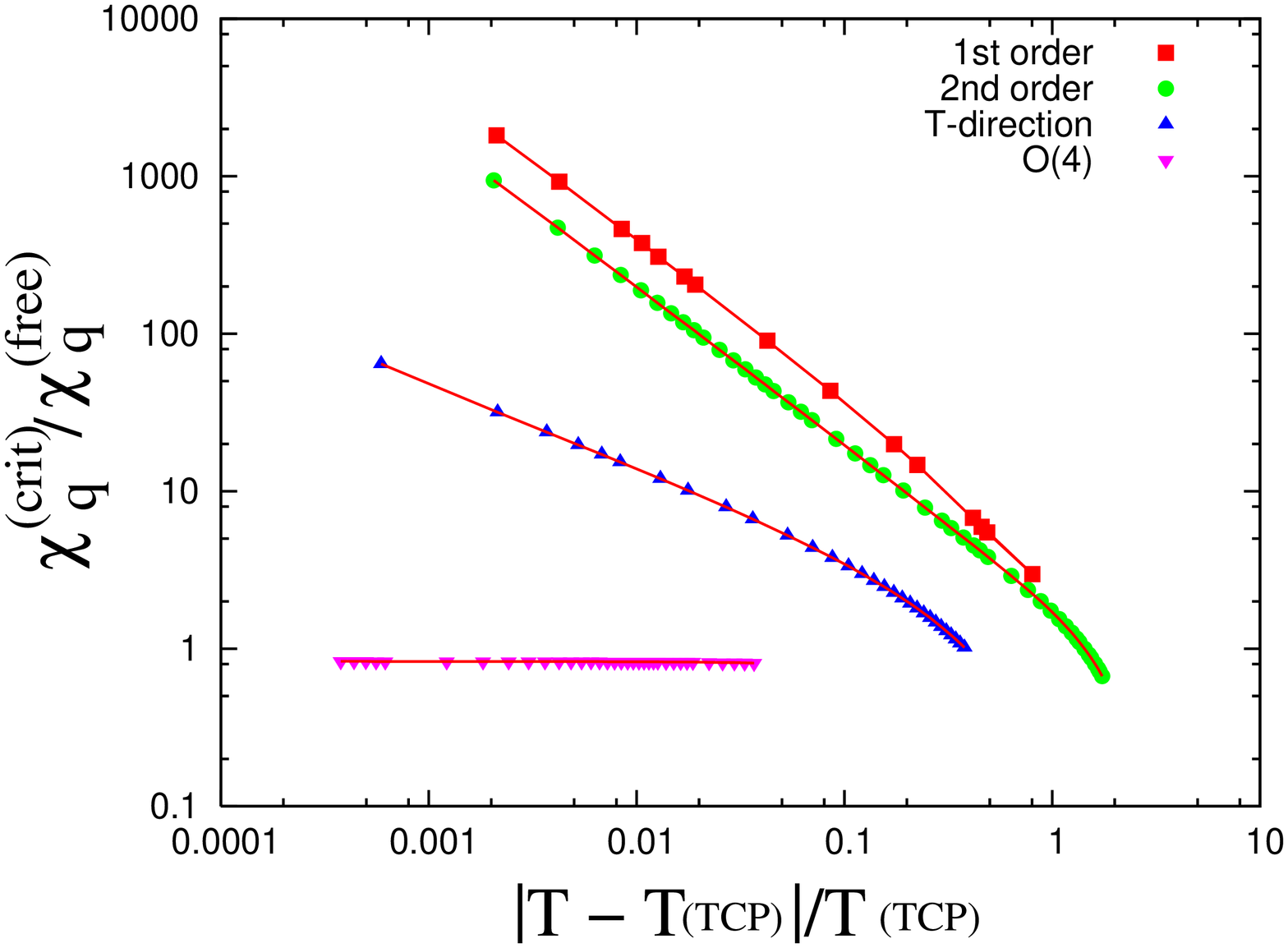}}
\begin{center}\vskip -0.3cm
\caption{\protect\label{crit-exp} The left-hand figure: the isovector fluctuations versus
$T$ for fixed $\mu_q$ chosen as  in the Fig. 2. The right-hand figure: the quark number
susceptibility $\chi_q$ near TCP versus $t=|T-T_c|/T_c$ when approaching the TCP along
the line of the 1st and the 2nd order transition  as well as at constant  $T$. Also shown
is $\chi_q$ when crossing  the O(4) critical line at fixed $\mu_q$ \protect{\cite{our}}.
} \vspace*{0.5cm} \end{center}
\end{figure}

 The existence of the
tricritical point (TCP) is seen in Fig. 2-left to be   strictly related with the strength
of the four-fermion interactions in the isovector-vector channel controlled by the
coupling $G_V$ in the NJL Lagrangian. For large  $G_V>0.6G_S$ (with $G_S$ describing the
strength of the scalar four-fermion interactions)  the TCP disappears from the phase
diagram due to strong repulsive interactions in the medium.
\begin{figure}\hskip 0.5cm
{\includegraphics[width=6.4cm,height=6.39cm]{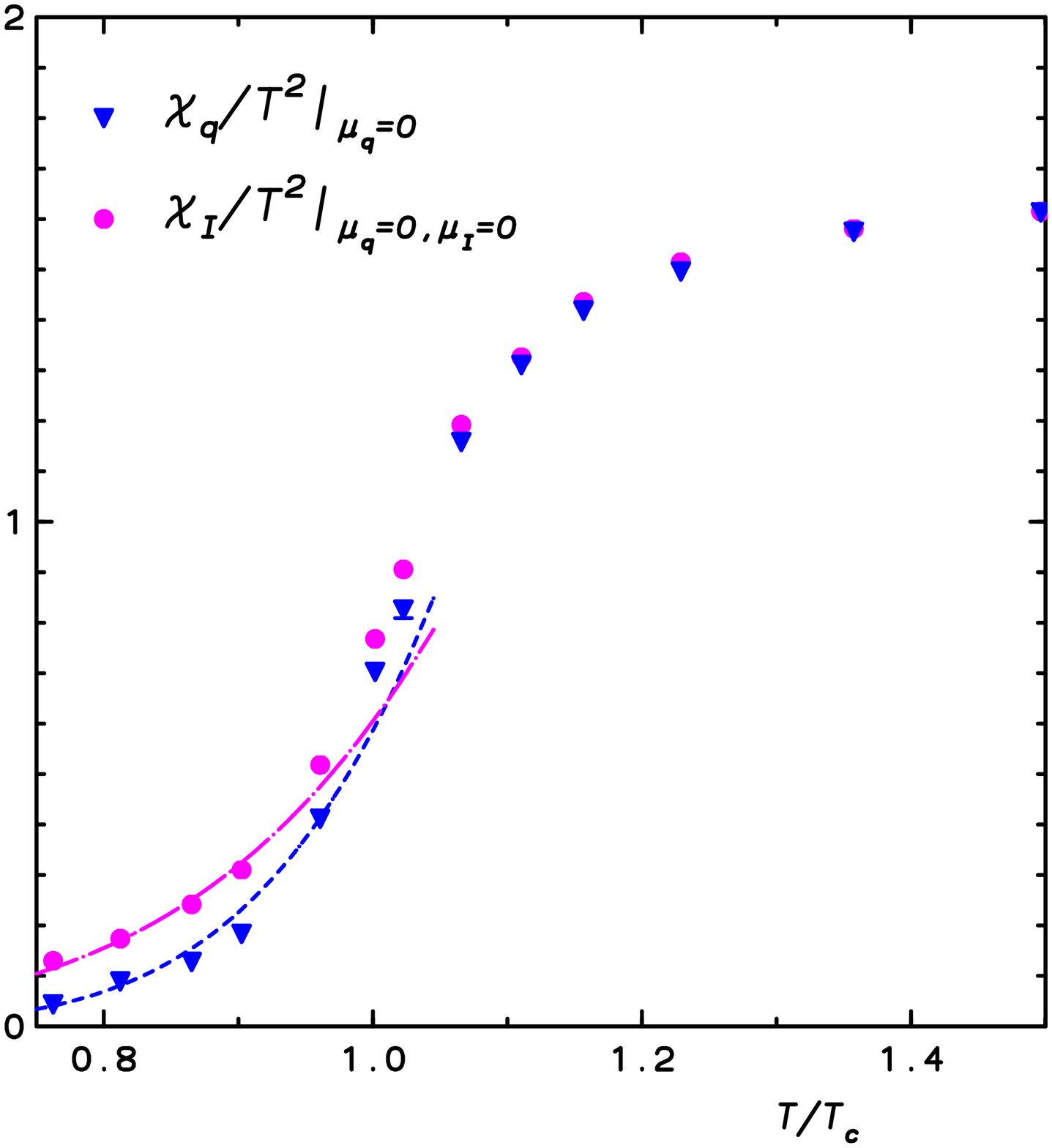}} {\vskip -13.9pc\hskip
19.3pc\includegraphics[width=6.5cm,height=4.9cm]{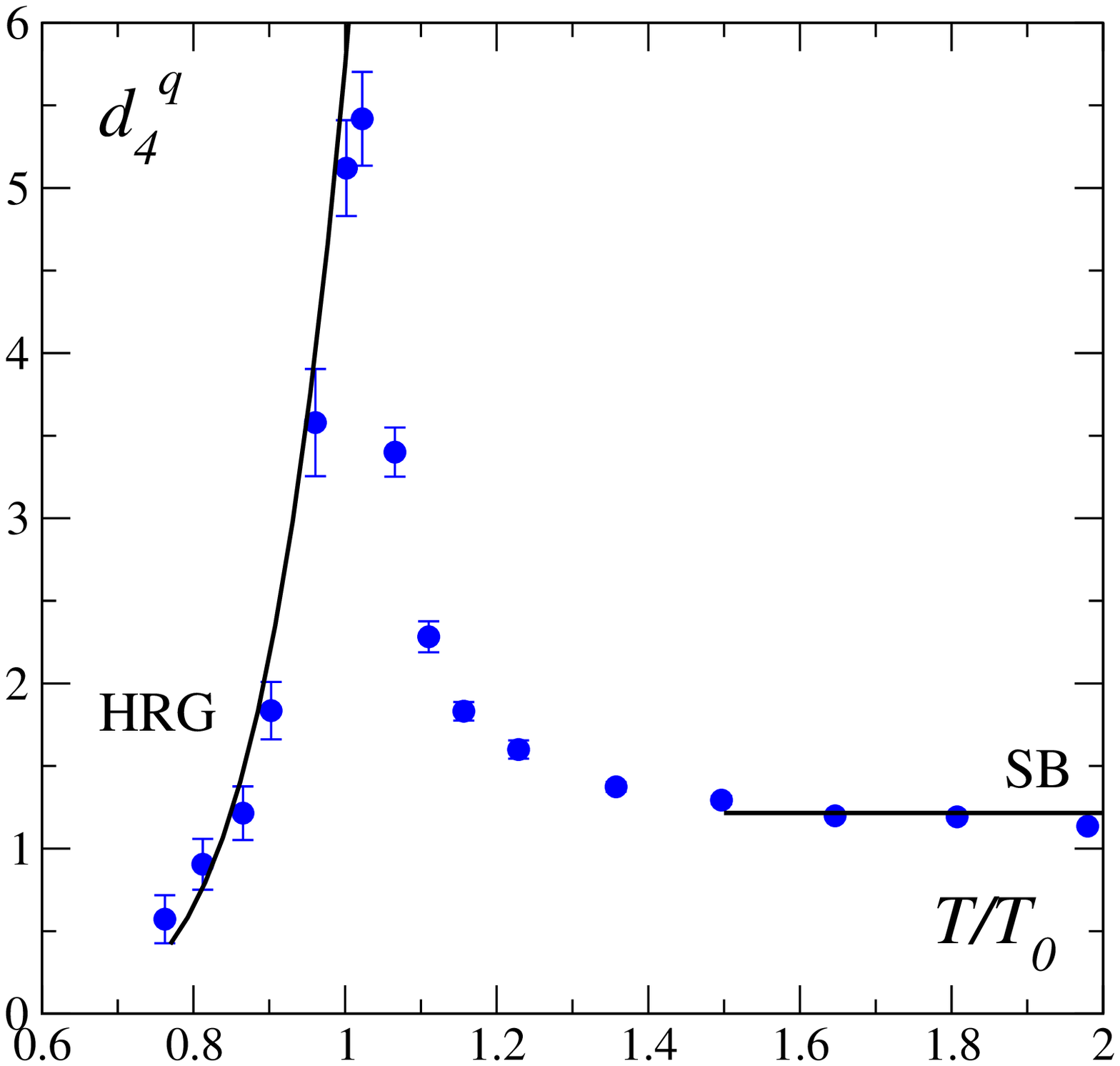}
 }\begin{center}\vskip -0.1pc\caption{\label{fig2:phase}
  The
left-hand figure: the isovector $\chi_I$ and the quark number $\chi_q$ susceptibilities
at $\mu_q=0$.
  The right-hand figure: the fourth-order cumulant moments. The
  results are from LGT calculations with $N_f=2$ and $m_q\simeq 80$ MeV \protect{\cite{LGT1,ejiri}}.
  The lines for $T/T_0<1$ are the hadron resonance gas model results.}
\end{center}
\vspace*{0.5cm}
\end{figure}

The position of   the TCP   can be identified by considering the fluctuations  of the
charge density $n_q$  that are quantified by the corresponding susceptibilitie
$\chi_q=\partial n_q/\partial\mu_q$ \cite{our,bj,hatta1}. Fig. 2-right shows the
$T$-dependence of the net quark number susceptibilities in the  vicinity and at the TCP.
The $\chi_q$ varies  rapidly with $T$ and drops discontinuously at the phase boundary.
The size of this discontinuity grows with increasing $\mu_q$ up to the TCP where the
susceptibility diverges. Beyond the TCP the discontinuity is again finite. At $\mu_q=0$
the discontinuity vanishes and the susceptibility shows a weaker, non-analytic structure
at the transition temperature. Also at finite quark mass the discontinuity along the
cross-over line and at the critical end point  is replaced by a peak which diverges at
the critical end point. The critical properties of $\chi_q$ at the TCP and along the 2nd
order line are universal and are governed by different critical exponents.

Fig. 3-right  illustrates the critical behavior of $\chi_q$  near the 2nd order  critical
line and at the TCP. In the mean field  approximation the critical dependence  of
$\chi_q$ on the reduced temperature $t=|T-T_c|/T_c$ is consistent with that  obtained  in
the Landau theory \cite{hatta,our}. For  paths approaching the TCP asymptotically
tangential to the phase boundary, the susceptibility diverges with the correlation length
critical exponent $\gamma_q=1$. For other paths the critical exponent is $\gamma_q= 1/2$.
Along the second order critical line, the susceptibility remains finite. Going beyond the
mean field and including quantum fluctuations in the lagrangian the above critical
exponents are renormalized to those  belonging to  the  universality class  of the 3D
Ising model \cite{bj}. On the other hand, the finite value of  $\chi_q$ along the 2nd
order line is consistent with the O(4) universality class \cite{our,ejiri,bj,hatta1}.
There, the critical properties of $\chi_q$ are governed by the specific heat critical
exponents $\alpha\simeq -0.2$ \cite{hatta,ejiri}.\footnote{In the NJL model we obtain the
mean-field value for this critical exponent, $\alpha=0$.}
 Consequently, at $\mu_q=0$  the
$T$-dependence of $\chi_q$ is only due to the regular part of the free energy
\cite{hatta,ejiri}. At finite $\mu_q$ the singular  contribution to $\chi_q$ results in
the cusp structure the strength of  which increases with  $\mu_q$. The fourth-order
cumulant $d^4_q\simeq
\partial^4 P/\partial\mu_q^4$ develops a cusp already at $\mu_q=0$ and diverges at finite
$\mu_q$ along the O(4)-line \cite{ejiri}.

Contrary to $\chi_q$, the isovector fluctuations $\chi_I$ shown in  Fig. 3-left  are
neither singular nor discontinuous at the chiral phase transition. The non-singular
behavior  of $\chi_I$ at the TCP is consistent with the observation that there is no
mixing between isovector excitations and the isoscalar sigma field due to the SU(2)$_V$
isospin symmetry\cite{hatta1}.

\begin{figure}
%
\hskip 0.5cm {\includegraphics[width=6.4cm,height=5.7cm]{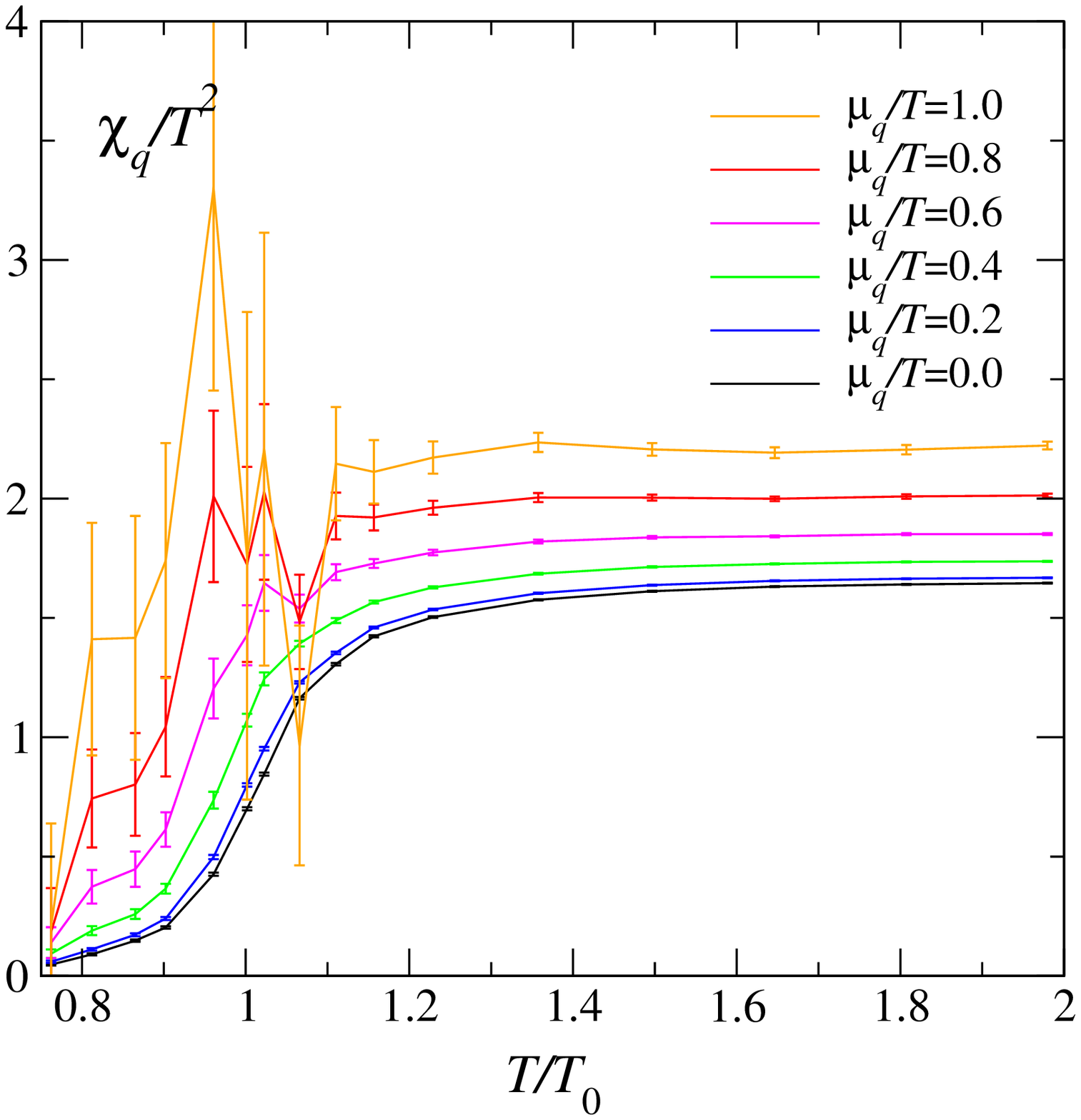}} {\vskip
-13.4pc\hskip 19.3pc\includegraphics[width=6.5cm,height=5.2cm]{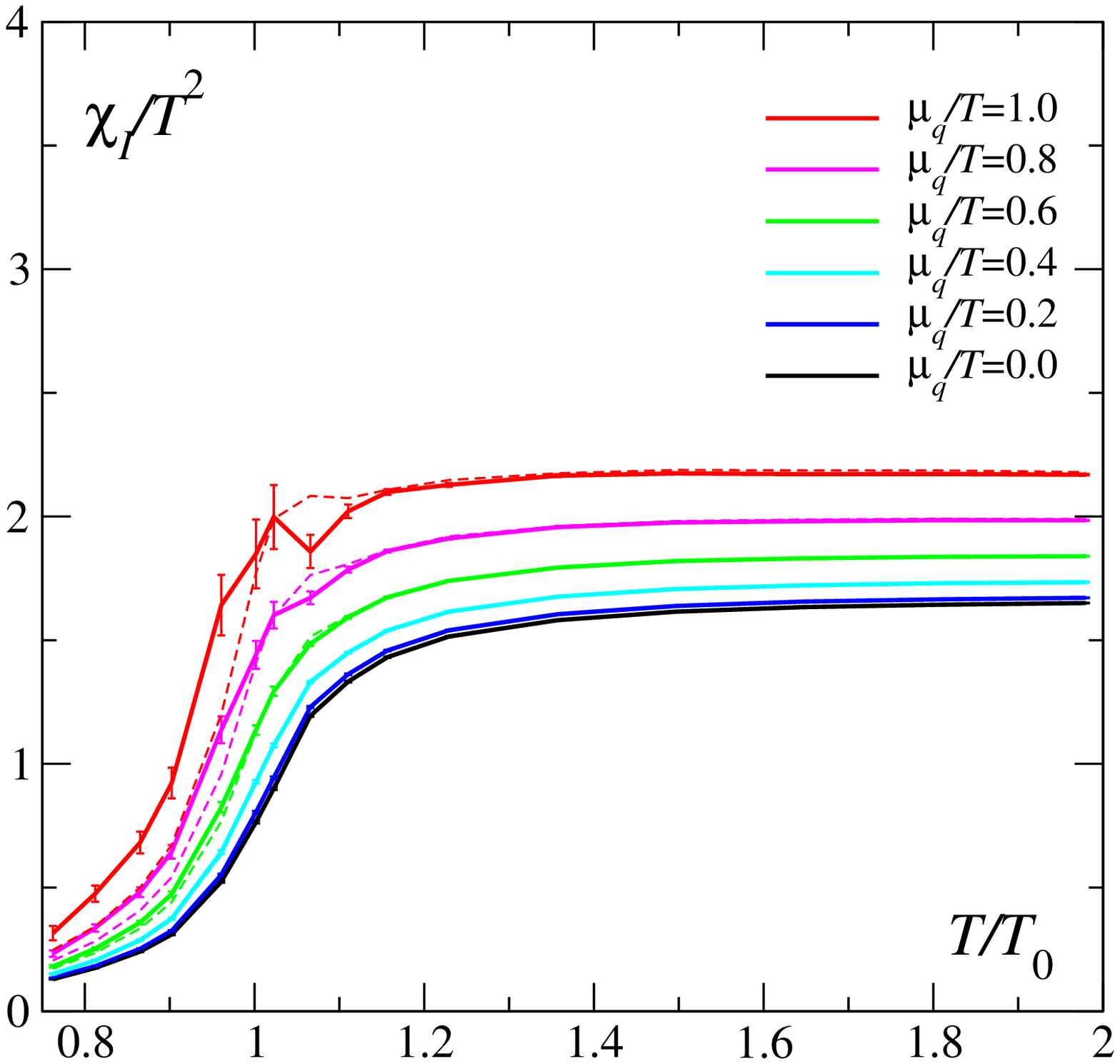}
 }
 \begin{center}\vskip -0.1pc\caption{\label{fig3:phase}
The  net quark number $\chi_q$  (left-hand figure) and isovector  $\chi_I$ (right-hand
figure) susceptibilities for different values of  $\mu_q/T$.  The results are from LGT
calculations with $N_f=2$ and $m_q\simeq 80$ MeV \protect{\cite{LGT1}}.  }
\end{center}
\vspace*{0.5cm}
\end{figure}

Recently,  the net    quark as well as  isovector fluctuations and higher moments  were
computed \cite{LGT1,ejiri} in two flavor lattice QCD (see Figs. 4 and 5). At $\mu_q=0$,
$\chi_q$
 shows a strong suppression of the fluctuations in confined phase and a cusp-like
structure   that increases with $\mu_q$ in the  vicinity of the transition temperature at
$\mu_q\neq 0$. Also observed on the lattice was a cusp in the forth-order cumulant moment
at $\mu_q=0$.   The lattice results confirmed that the quark fluctuations in the
isovector channel $\chi_I$, contrary to $\chi_q$, do not show any peak structure and
depend   only weakly on $\mu_q$. These  properties of $\chi_q$ and $\chi_I$ are
consistent with O(4) universality arguments and should be there when approaching the
critical end point with increasing $\mu_q$. However, the above behavior of $\chi_q$ and
$\chi_I$ can be also interpreted in terms of  the regular part of the free energy due to
the enhanced contribution of resonances in the  vicinity of the transition temperature.
The LGT results on $T$ and $\mu_q$ dependence of  $\chi_q$ and $\chi_I$ as well as
$d_q^4$  can  in the confined phase  be  successfully described \cite{ejiri}  by the
resonance gas partition function  as seen in Figs. 4. This is also illustrated  in  Fig.
6, where the ratio $R_{4,2}^q=d_4^q/d^2_q$ of the fourth- to the second-order cumulants
as well as the ratio of the quark density to the susceptibility are seen to be consistent
with the hadron resonance gas  model. The observed change in cumulants  at $T_c$ is due
to deconfinement resulting in a change of the quark content of particles in a medium.
Consequently, the properties of the charge fluctuations seen in Figs. 4-6 are only
necessary but not sufficient conditions to verify the existence of a critical endpoint
(CEP) in the phase diagram. To verify  the existence of the  CEP one would need to
observe  a non-monotonic behavior of the net quark number (or electric charge)
susceptibilities as shown in  Fig. 7. The non-monotonic behavior of $\chi_q$ along the
boundary line appears since the quark number fluctuations are finite along the 2nd  and
the 1st order transition and  diverge at the TCP.

The finite structure of the charge susceptibilities along the first order line is valid
under the assumption that this transition appears in equilibrium. Admitting deviations
from the idealized equilibrium situation, the first order phase transition is intimately
linked with the existence of a convex anomaly in thermodynamic pressure \cite{ran}. There
is an interval of energy or baryon number density where the derivative of pressure
$\partial P/{\partial V}>0$. Such an anomalous behavior describes the region of
instabilities bounded  by the spinodal lines where the pressure derivatives taken at
constant temperature or entropy vanishes. 
Changing the temperature or entropy results as the isothermal or isentropic spinodal line
in the ($T,n_q)$-plane with $n_q$ being the net quark number density. 
 The spinodal decomposition of the system was argued to enhance the baryon number  and strangeness
fluctuations \cite{ran}.

\begin{figure}
%
\hskip 0.5cm {\includegraphics[width=6.4cm,height=5.5cm]{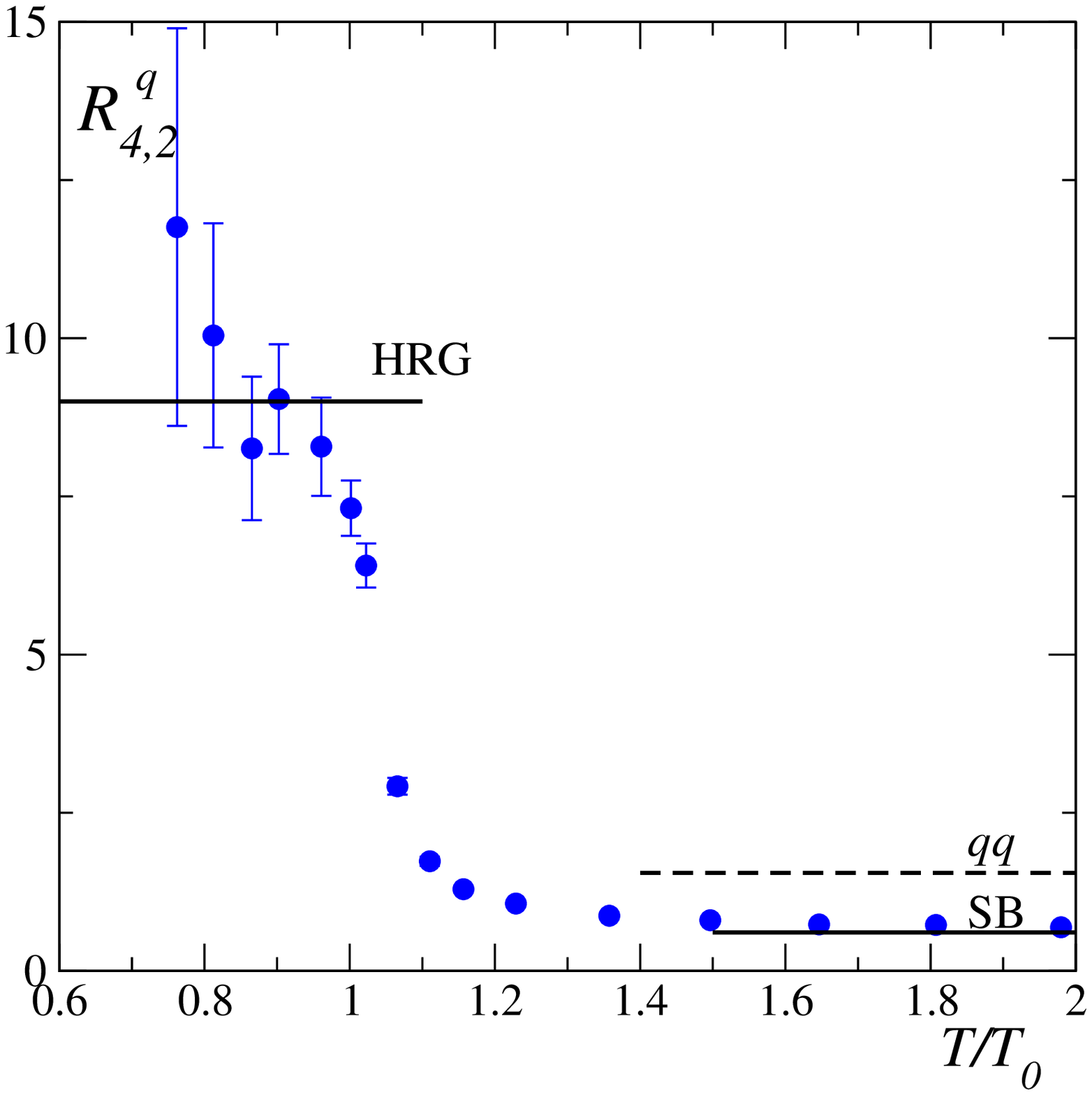}} {\vskip -13.0pc\hskip
19.3pc\includegraphics[width=6.5cm,height=5.1cm]{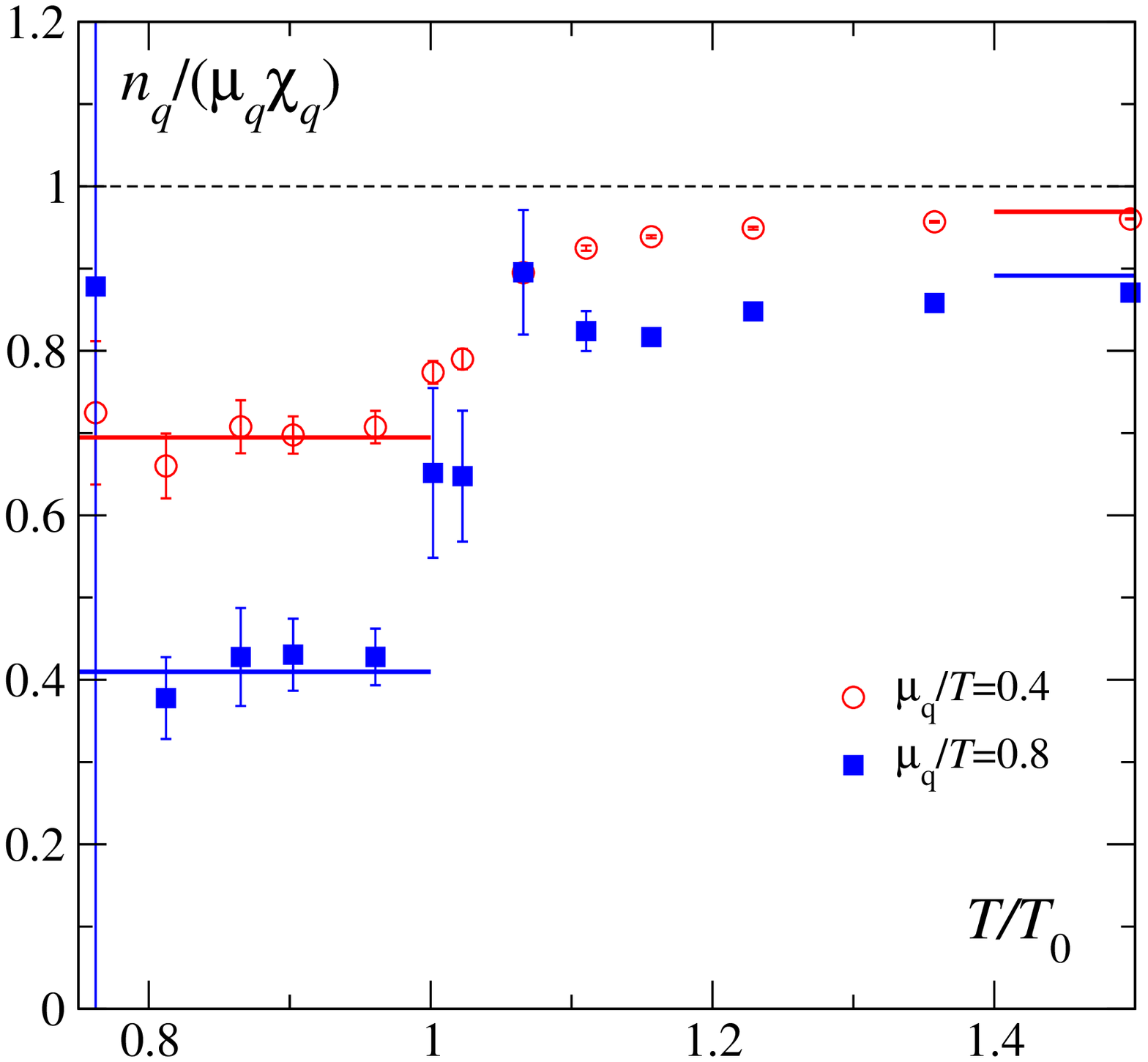}
 }
 \begin{center}\vskip -0.1pc\caption{\label{fig4:phase}
The ratio of fourth- to second-order cumulants of quark number at $\mu_q=0$ (left-hand
figure). The quark number density $n_q$  with respect to the net quark number
fluctuations $\chi_q$ (right) as a function of  $T/T_0$ for various $\mu_q/T$. Horizontal
lines show the infinite temperature ideal gas values and the HRG model prediction (solid
lines). The LGT results are from \protect{\cite{LGT1,ejiri}}
   }
\end{center}
\vspace*{0.5cm}
\end{figure}

Fig. 8-left  shows the phase diagram obtained in the NJL model formulated at finite
current quark mass  that accounts for  spinodal instability. From this figure   it is
clear that the isothermal spinodal lines end up at the CEP. It is thus natural to explore
how the charge fluctuations develop when going beyond the critical end point in the
direction of the first order phase transition.
 Fig. 8-right  shows the evolution of the net quark number fluctuations along
the path of fixed $T=50$ MeV  in the $(T,n_q)$--plane. The resulting behavior is quite
interesting. When entering the mixed phase  there is singularity   in  $\chi_{q}$ that
appears when crossing the isothermal spinodal lines, where  the fluctuations diverge and
change the sign.
Between the spinodal lines  the susceptibility  is negative. This implies a strong
instability in the baryon number fluctuations when crossing the transition region between
chirally symmetric and broken phase.

\begin{figure}
\begin{center}
\includegraphics[width=7.1cm]{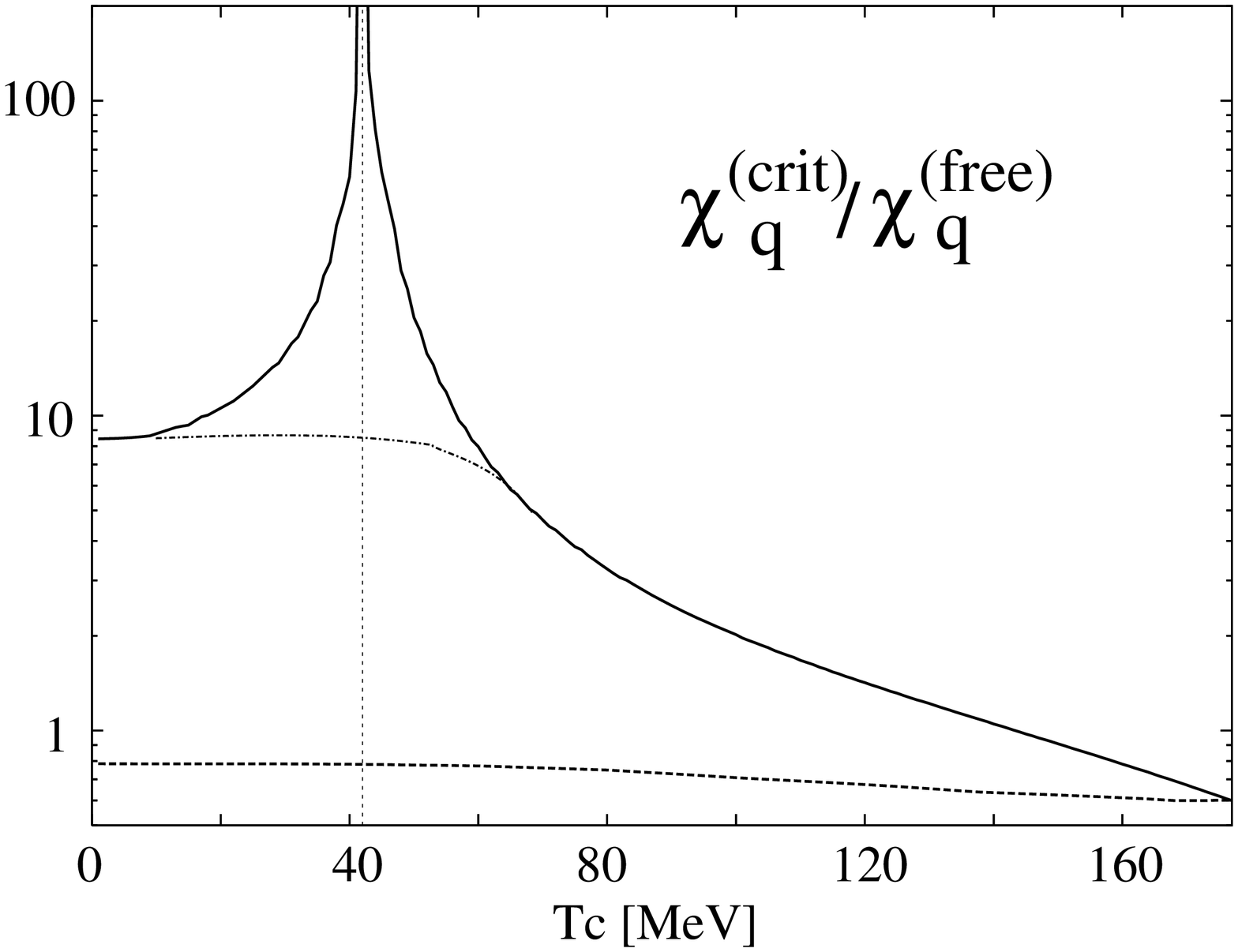}
\includegraphics[width=7.1cm]{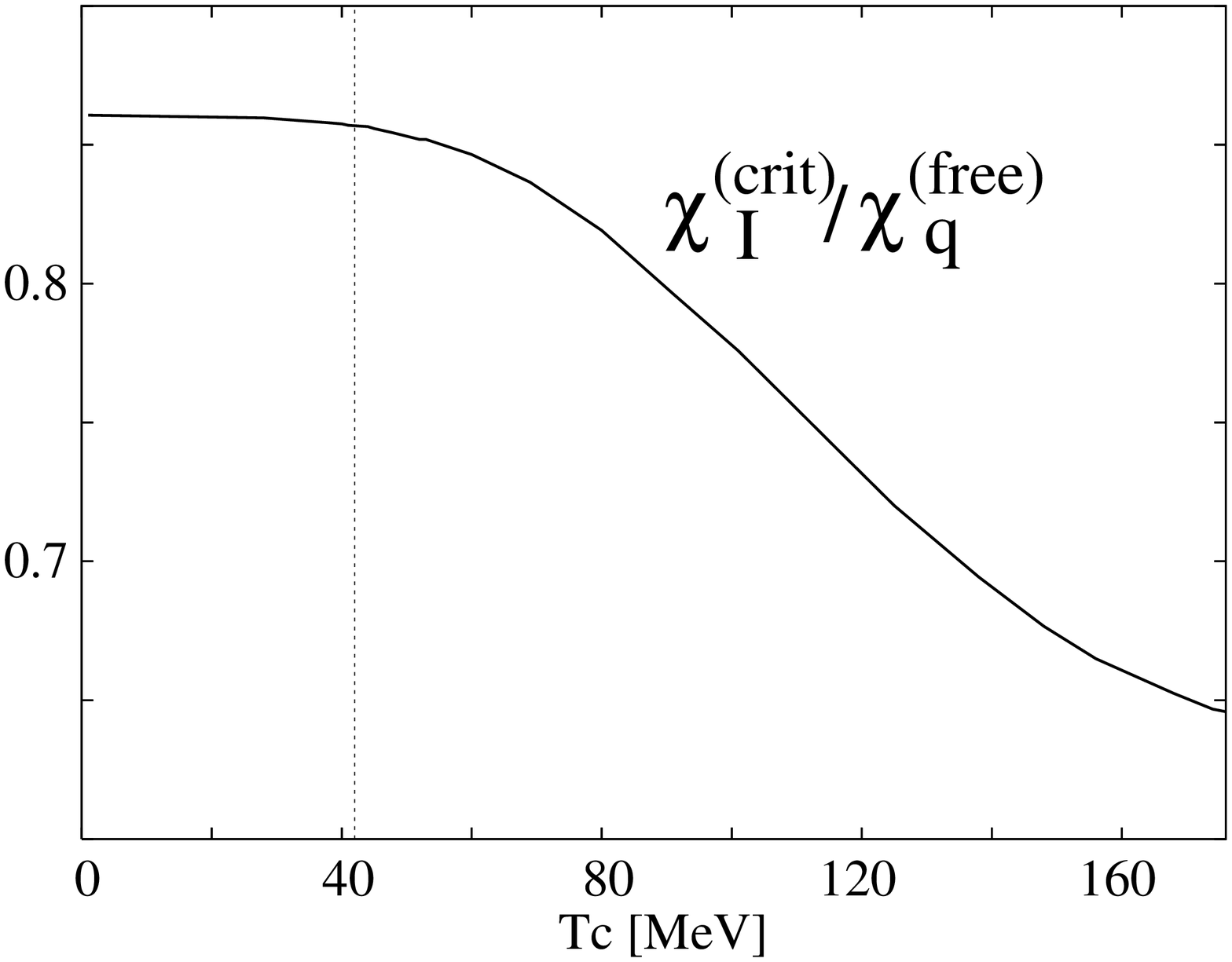}
\end{center}
\caption{\protect\label{tcp_q} The quark number (left) and isovector (right)
susceptibilities $\chi_q$ and $\chi_I$ as functions of the temperature along the phase
boundary. In the left--hand figure the solid (dashed) line denotes $\chi_q$ in the
chirally broken (symmetric) phase. The vertical dotted-line indicates the position of the
tricritical point TCP. The calculations were done in the chiral limit in the NJL model
\protect{\cite{our}}.}
\vspace*{0.5cm}
\end{figure}


The behavior of $\chi_{q}$ seen in Fig. 8 is a direct consequence of the following
thermodynamic relation
\begin{equation}\label{eq7}
\left( {{\partial P}\over {{\partial V}}} \right)_T = - {{n_q^2}\over {V}}{{1}\over
{\chi_{q}}}\,,
\end{equation}
what connects  the pressure derivative with the  charge density $n_q$ and the
corresponding susceptibility $\chi_{q}$. Along the isothermal spinodal lines the pressure
derivative in (\ref{eq7}) vanishes. Thus, for non vanishing density $n_q$ its
fluctuations have to diverge to satisfy  the relation (\ref{eq7}). In addition, since the
pressure derivative ${\partial P}/{\partial V}|_T$ changes   sign when crossing the
spinodal points, there should be discontinuity and the corresponding sign change in the
divergence of $\chi_{q}$ as seen in Fig. 8.
Due to linear relation between $\chi_q$, $\chi_I$ and the electric charge susceptibility
$\chi_Q$  a similar behavior as seen  in Fig. 8  for $\chi_{q}$  should be also there for
$\chi_Q$.

From    Fig. 8 it is clear that when approaching the  CEP from the side of the first
order transition the region of instability shrinks and  disappears completely at the CEP.
Consequently, the CEP singularity in $\chi_{q}$ appears from the matching of the two
positive singular branches of $\chi_{q}$. Thus, the singular properties at the CEP are
the remnant of divergent behavior of $\chi_{q}$ at the first order chiral phase
transition.
From the above discussion it  is clear that divergent charge fluctuations  are not only
attributed to the second order critical end point or TCP  but are also there at the first
order transition if the spinodal phase decomposition sets in. Consequently, the
fluctuations of the conserved charges  could  be considered as a signal of the first
order transition that is expected in the QCD phase diagram~\cite{sfr}.
{\begin{figure}
\begin{center}
\includegraphics[width=7.4cm]{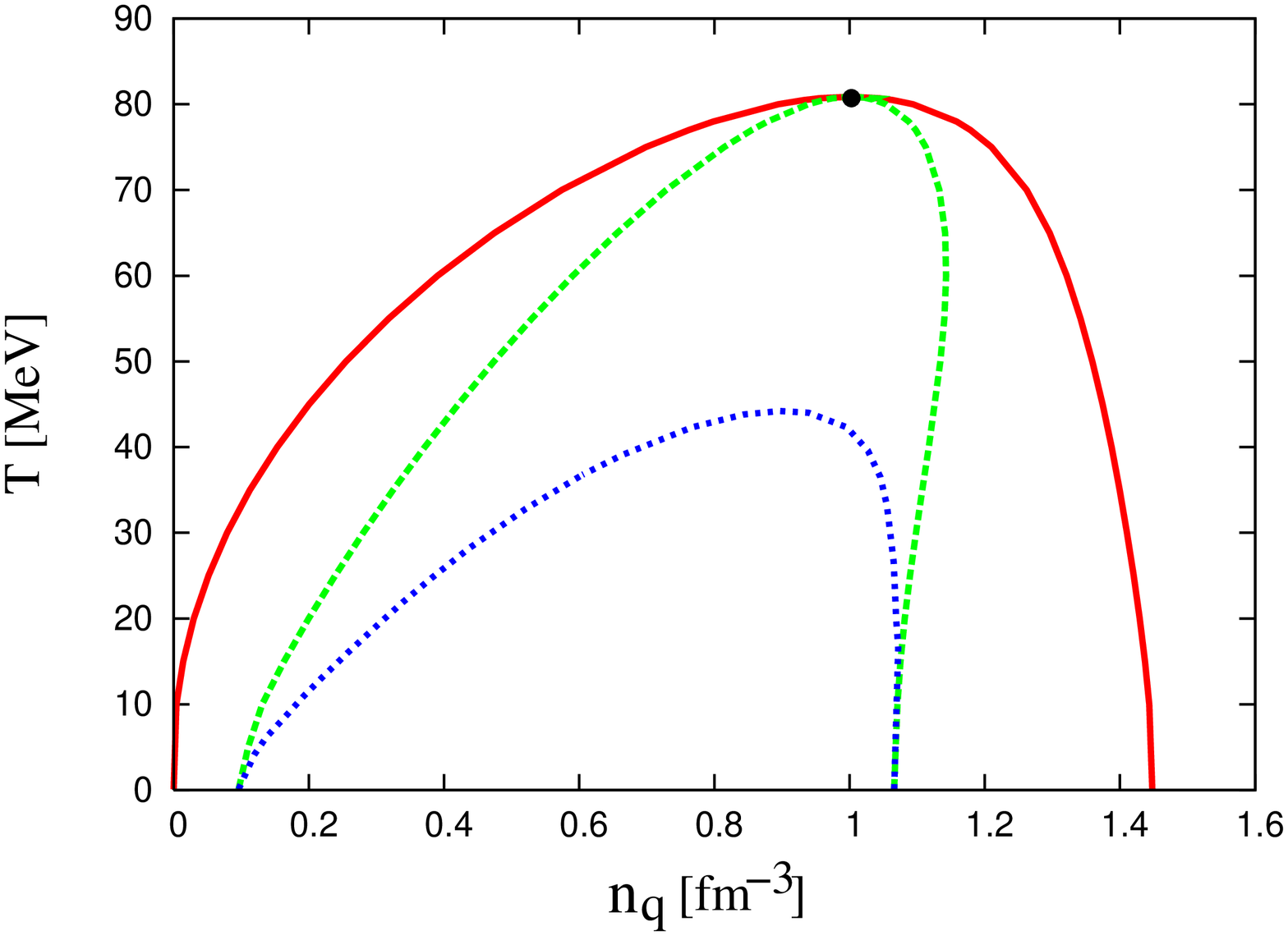}
\includegraphics[width=7.4cm]{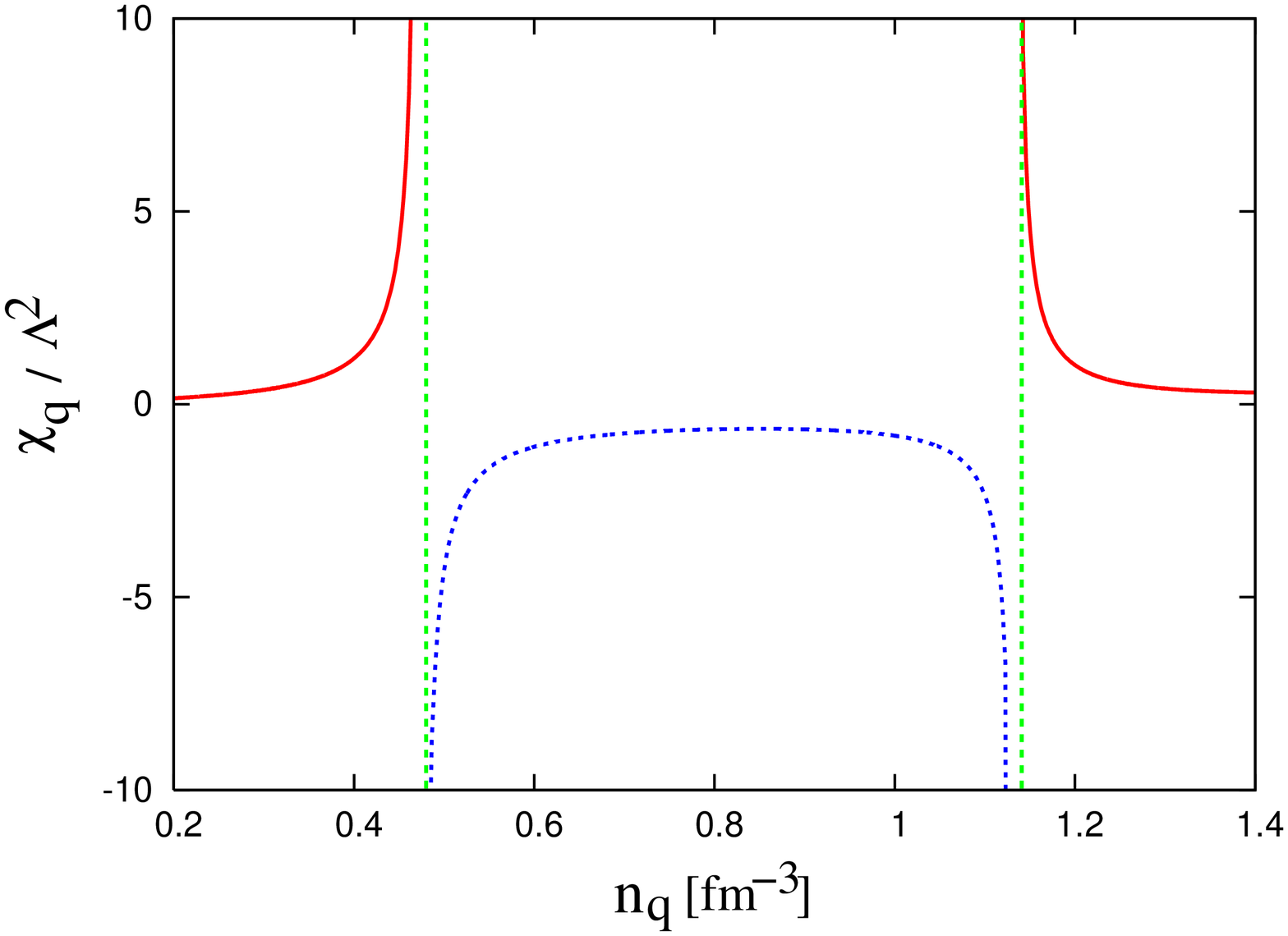}
\caption{\label{fig5:phase} The left-hand figure: The phase diagram in the temperature
$T$ and quark number  density $n_q$ plane
 in the NJL model. The filled point indicates the CEP.
The full lines starting at the  CEP represent  boundary of the mixed phase in
equilibrium. The broken-curves are the isothermal whereas the dashed-broken ones are the
isentropic spinodal lines.
  The right-hand figure: the  net quark
number density fluctuations $\chi_{q}/\Lambda^2$ normalized to momentum cutt-off
$\Lambda$ as a function of the quark  number density $n_q$ across the first order phase
transition. The $\chi_{q}$ was calculated in the NJL model along the line of constant
temperature $T=50$ MeV.}
\end{center}
\vspace*{0.5cm}
\end{figure}
 }

 In summary, we have presented a brief discussion of the phase structure
 expected in two flavor QCD.  We have discussed the properties of charge density
fluctuations when crossing the phase boundary. It was shown that for equilibrium
transition the fluctuations of the net baryon number and electric charge exhibit a
non-monotonic behavior along the QCD phase boundary  that could be applied  in heavy ion
collisions as the signal for the existence of the critical end point. We have also argued
that admitting deviations from the equilibrium picture for the first order transition the
above non-monotonic structure is to be modified. In the presence of spinodal
instabilities the fluctuations of the conserved charges diverge when crossing the
isothermal spinodal lines of the 1st order transition. Consequently, such fluctuations
can  be used  to probe   the first order phase transition in heavy ion collisions.
\newpage

K.R. acknowledges fruitful  collaboration and discussions with S. Ejiri and  F. Karsch
and the suport of the Polish Ministry of National Education (MEN).

\end{document}